\documentstyle[sprocl,epsfig,here,rotating]{article}

\def\a{\alpha}
\def\g{\gamma}
\def\d{\delta}
\def\D{\Delta}
\def\h{\eta}

\def\O{\Theta}
\def\p{\pi}
\def\r{\rho}
\def\f{\phi}
\def\W{\Omega}

\def\ol{\overline}
\def\hs{\hskip}
\def\vs{\vskip}
\def\\{\hfill\break}

\def\ran{\rangle}
\def\lan{\langle}
\def\hf{\hfill}

\def\aver#1{\langle#1\rangle}
\def\XVEC#1#2{#1_1,\ldots,{#1}_{#2}}

\def\FACT#1#2{{#1}_m({#1}_m-1)\ldots({#1}_m-#2+1)}
\def\ifmath#1{\relax\ifmmode #1\else $#1$\fi}%

\newcommand{\beq}{\begin{equation}}
\newcommand{\eeq}{\end{equation}  }
\newcommand{\beqa}{\begin{eqnarray}}
\newcommand{\eeqa}{\end{eqnarray}  }

\def\ifmath#1{\relax\ifmmode #1\else $#1$\fi}%
\def\rd{\ifmath{{\mathrm{d}}}}
\def\rL{\ifmath{{\mathrm{L}}}}

\def\rS{\ifmath{{\mathrm{S}}}}

\def\all{\ifmath{{\mathrm{all}}}}
\def\vec#1{{\mbox{\bf #1}}}
\def\lan{\langle}
\def\ran{\rangle}

\begin{document}

\bibliographystyle{unsrt}
~\vs -1.5cm
\hf Nijmegen preprint

\hf HEN-400

\hf December 1996
\vs 5mm
\title{CORRELATIONS AND FLUCTUATIONS IN HIGH-ENERGY COLLISIONS
\footnote{Invited review given at the Workshop 
``The Status of Physics at the End of the 20th Century'', Santa Fe (U.S.A.), 
October 1996}}
\author{W. KITTEL}

\address{HEFIN, University of Nijmegen,\\ 
Toernooiveld 1, 6525 ED Nijmegen, The Netherlands}

\maketitle
\abstracts{
In addition to its importance in describing high-energy processes themselves,
the dynamics of multiparticle production is part of the general field of 
non-linear phenomena and complex systems. Multiparticle dynamics is one 
of the rare fields of physics where {\it higher}-order correlations are 
directly accessible in their full multi-dimensional characteristics under 
well-controlled experimental conditions. Multiparticle dynamics, therefore,
is an ideal testing ground for the development of advanced statistical 
methods.\\
Higher-order correlations have, indeed, been observed as particle-density 
fluctuations. Approximate scaling with finer resolution provides evidence 
for a self-similar correlation effect. Quantum-Chromodynamics branching is a 
good candidate for a dynamical explanation of these correlations in 
e$^+$e$^-$ collisions at CERN/LEP and, as expected, also of those in pp
collisions at future CERN/LHC energies. However, other sources such as 
identical-particle Bose-Einstein interference effects also contribute. \\
A particular question at the moment is the smooth transition from the QCD 
branching domain (gluon interference before hadronization) to the 
Bose-Einstein domain (identical-pion interference after 
hadronization). Both mechanisms have clearly been observed in e$^+$e$^-$ 
collisions at CERN/LEP energies. The large amount of high-resolution data 
being collected at LEP will allow the study of the 
{\it genuine} (i.e. non-trivial) higher-order correlations in both domains.}

\section{Introduction}
Recent years have witnessed a remarkable amount of experimental and 
theoretical activity in search of scale-invariance and fractality in 
multihadron production processes. In addition to being an important part 
of high energy physics itself, the dynamics of multi-particle production 
in collisions of elementary particles at high energies (multiparticle 
dynamics) is part of the general field of non-linear phenomena and
complex systems. Studies of classical and quantum chaos, non-equilibrium 
dissipative processes, random media, growth phenomena and many more 
processes have all contributed in revealing the pervasive importance
of self-similarity, power-laws, and fractals in nature. Research in these 
fields is still in full swing and continues to uncover intriguingly simple 
and often surprisingly universal behavior in complex, non-linear systems. 

While considerable experience already exists in many fields for the study 
of two-component correlation, it is often in {\it higher}-order (i.e. 
multi-component) correlations that the most interesting properties manifest
themselves, in the simultaneous interplay of a large number of components. 
The special significance of multiparticle dynamics for the development of 
advanced statistical methods lies in the fact that it is one of the rare 
fields of physics where higher-order correlations are directly accessible 
in their full multi-dimensional characteristics under well controlled 
experimental conditions.

Higher-order correlations have recently been observed as particle-density 
fluctuations in cosmic ray, nucleus-nucleus, hadron-hadron, e$^+$e$^-$ and 
lepton-hadron experiments. To study these fluctuations in detail, normalized 
correlation integrals are being analyzed in phase-space domains of ever 
decreasing size. Approximate scaling with decreasing domain size is now 
observed in all types of collision, giving evidence for a correlation effect 
self-similar over a large range of the resolution (called intermittency, 
in analogy to a statistically similar problem in spatio-temporal turbulence). 

Parton branching of Quantum-Chromodynamics predicts the type of correlations 
observed in e$^+$e$^-$ collisions at CERN/LEP and, as expected, also in pp
collisions at CERN/LHC energies. However, other sources such as 
Bose-Einstein interference of identical particles also contribute. Fast 
development of the applied technology has taken place over the last few 
years, in particular in the extension of originally one-dimensional to 
full three-dimensional phase space analysis.

\section{Methodology}
\subsection{Particle Densities}

A collision between particles a and b is assumed to yield
exactly $n$ particles in a sub-volume $\Omega$ of total phase space
$\Omega_{\all}$. The single symbol $y$ represents the 
kinematical variables needed to specify the position of each particle in 
this space (for example, $y$ can be the full four-momentum of a particle
and $\W$ a cell in invariant phase space or simply the c.m. 
rapidity~\footnote{\ Rapidity $y$ is defined as
$y={1\over 2} \ln [(E+p_\rL)/(E-p_\rL)]$, with $E$ the energy and $p_\rL$ the
longitudinal component of momentum vector $\vec p$ along a given direction
(beam-particles, jet-axis, etc.); pseudo-rapidity is defined as
$\h={1\over 2}\ln [(p+p_\rL)/(p-p_\rL)]$.}
of a particle and $\Omega$ an interval of length $\delta y$).  
The distribution of points in $\Omega$ can then be characterized
by continuous probability densities $P_n(y_1,\ldots,y_n)$; $n=1,2,\ldots$.
For simplicity, we assume all  final-state particles to be of the same type.
In this case, the {\bf exclusive} distributions $P_n(y_1,\ldots,y_n)$
can be taken as fully symmetric in
$y_1,\ldots,y_n$; they describe the distribution in $\Omega$ when the
multiplicity is exactly $n$.

The corresponding {\bf inclusive} distributions are given for
$n=1,2,\ldots$ by:
\begin{eqnarray}
\rho_n(y_1,\ldots,y_n) &=& P_n(y_1,\ldots,y_n)\nonumber\\
& & \hskip-2cm +\sum_{m=1}^{\infty}\frac{1}{m!} \int_\Omega
P_{n+m}(y_1,\ldots,y_n,y'_1,\ldots,y'_m)\,\prod^m_{i=1} \rd y'_{i}\ .
\label{dr:1}
\end{eqnarray}
The inverse formula is
\begin{eqnarray}
P_n(y_1,\ldots,y_n)&=&\rho_n(y_1,\ldots,y_n)\nonumber\\
& & \hskip-2cm \mbox{}+\sum_{m=1}^{\infty}(-1)^m \frac{1}{m!} \int_\Omega
\rho_{n+m}(y_1,\ldots,y_n,y'_1,\ldots,y'_m)\,\prod^m_{i=1} \rd y'_i\ .
\label{dr:2}
\end{eqnarray}
Here, $\rho_n(y_1,\ldots,y_n)$ is the probability density  for $n$ 
points to be at $\XVEC{y}{n}$, irrespective of the presence and location of 
any further points. Integration over an interval $\W$ in $y$ yields
\begin{eqnarray}
 &~&\int_\W \r_1(y) dy = \lan n\ran\ , \nonumber \\
 &~&\int_\W \int_\W \r_2(y_1,y_2)dy_1dy_2 =
\lan n(n-1)\ran \ ,\nonumber \\
 &~&\int_\W dy_1 \dots \int_\W dy_q \r_q (y_1,\dots,y_q) =
\lan n(n-1)\dots (n-q+1)\ran \ ,
\end{eqnarray}
where the angular brackets imply the average over the event ensemble.

\subsection{Cumulant Correlation Functions}

Besides the interparticle {\it correlations} we are looking for,
the inclusive $q$-particle densities $\rho_q(\XVEC{y}{q})$ in general
contain ``trivial'' contributions from lower-order densities. 
It is, therefore, advantageous to consider a new sequence of functions
$C_q(\XVEC{y}{q})$ as those statistical quantities which vanish whenever one
of their arguments becomes statistically independent of the others.
Deviations of these functions from zero shall be addressed as {\it genuine}
correlations. 

The quantities with the desired properties are the correlation 
functions - also called (factorial) cumulant functions - or, in integrated 
form, Thiele's semi-invariants.\cite{thiele} A formal proof of this property 
was given by Kubo.\cite{kubo}
The cumulant correlation functions are defined as in the cluster expansion
familiar from statistical mechanics via the sequence:
\cite{kahn:uhlenbeck,huang,Mue71}
\begin{eqnarray}
\rho_1(1)& =& C_1(1),\\
\rho_2(1,2)& =& C_1(1)C_1(2) +C_2(1,2),\\
\rho_3(1,2,3)& =& C_1(1)C_1(2)C_1(3)
+C_1(1)C_2(2,3)
+C_1(2)C_2(1,3)
+\nonumber\\
& &\mbox{}
+C_1(3)C_2(1,2)+C_3(1,2,3);
\end{eqnarray}
and, in general, by
\begin{eqnarray}
\rho_m(1,\ldots,m) &=& \sum_{{\{l_i\}}_m}\sum_{\mbox{perm.}}
\underbrace{\left[C_1()\cdots C_1()\right]}_{l_1\,\mbox{factors}}
\underbrace{\left[C_2(,)\cdots C_2(,)\right]}_{l_2\,\mbox{factors}}
 \cdots\nonumber\\
& & \cdots \underbrace{\left[C_m(,\ldots,)\cdots C_m(,\ldots,)
\right]}_{l_m\,\mbox{factors}}.
\label{a:4}
\end{eqnarray}
Here, $l_i$ is either zero or a positive integer and the sets of integers
$\{l_i\}_m$ satisfy the condition
\begin{equation}
\sum_{i=1}^m i\, l_i=m.\label{a:5}
\end{equation}
The arguments in the $C_i$ functions are to be filled by the $m$ possible
momenta in any order. The sum over permutations is a sum over all
distinct ways of filling these arguments. For any given factor product there
are precisely~\cite{huang}
\begin{equation}
\frac{m!}{
\left[(1!)^{l_1} (2!)^{l_2}\cdots(m!)^{l_m}\right] {l_1!}{l_2!}\cdots{l_m!}}
\label{a:6}
\end{equation}
terms.
    
The relations (\ref{a:4}) may be inverted with the result:
\begin{eqnarray}
C_2(1,2)&=&\rho_2(1,2) -\rho_1(1)\rho_1(2)\ ,\nonumber\\
C_3(1,2,3)&=&\rho_3(1,2,3)
-\sum_{(3)}\rho_1(1)\rho_2(2,3)+2\rho_1(1)\rho_1(2)\rho_1(3)\ ,\nonumber\\
C_4(1,2,3,4)&=&\rho_4(1,2,3,4)
-\sum_{(4)}\rho_1(1)\rho_3(1,2,3)
-\sum_{(3)}\rho_2(1,2)\rho_2(3,4)\nonumber\\
&&\mbox{} +2\sum_{(6)}\rho_1(1)\rho_1(2)\rho_2(3,4)-6\rho_1(1)
\rho_1(2)\rho_1(3)\rho_1(4).
\label{a:4b}
\end{eqnarray}
In the above relations  we have abbreviated $C_q(\XVEC{y}{q})$ to
 $C_q(1,2,\ldots,q)$; the summations indicate that all possible permutations
must be taken (the number under the
summation sign indicates the number of  terms).
Expressions for higher orders can be derived from the related formulae given
in~\cite{kendall}.
 
It is often convenient to divide  the functions
$\rho_q$ and $C_q$ by the product of one-particle densities, which leads to
the  definition of the  normalized inclusive densities and correlations:
\begin{eqnarray}
r_q(\XVEC{y}{q}) &=& \rho_q(\XVEC{y}{q})/\rho_1(y_1)\ldots
\rho_1(y_q),\label{3.8}\\
K_q(y_1,\ldots,y_q)& =& C_q(y_1,\ldots,y_q)/\rho_1(y_1)\ldots
\rho_1(y_q).\label{3.9}
\end{eqnarray}
In terms of these functions, correlations have been studied extensively for 
$q=2$. Results also exist for $q=3$, but usually the statistics (i.e. number 
of events available for analysis) are too small to isolate genuine 
correlations. To be able to do that for $q\geq 3$, one must apply moments 
defined via the integrals in Eq.~3, but in limited phase-space cells.

\subsection{Cell-Averaged Factorial Moments and Cumulants}

In practical work, with limited statistics, it is almost always
 necessary to perform averages over more than a single phase-space
 cell.  Let $\Omega_m$ be such a cell (e.g. a single rapidity interval
 of size $\delta y$) and divide the phase-space volume into $M$
 non-overlapping cells $\Omega_m$ of size $\d\Omega$ independent of
 $m$.  Let $n_m$ be the number of particles in cell $\Omega_m$.
 Different cell-averaged moments may be considered, depending on the
 type of averaging.
 
Normalized cell-averaged factorial moments~\cite{bialas} are defined as
\begin{eqnarray}
F_q(\delta y)&\equiv&
\frac{1}{M}\;\sum_{m=1}^M \frac{\aver{\FACT{n}{q}}}{\aver{n_m}^q}
\label{dr:44}\\
&\equiv & \frac{1}{M} \sum^M_{m=1}
\frac{\int_{\delta y} \rho_q(y_1,\ldots,y_q) \prod^q_{i=1} dy_i}
{\left(\int_{\d y} \r(y)dy\right)^q}\ \   \\
&= & \frac{1}{M(\d y)^q} \sum^M_{m=1} \int_{\d y} \frac{\r_q(y_1,\dots,y_q)
\prod^q_{i=1} dy_i}{\left(\bar\r_m\right)^q}. \label{dr:45}
\end{eqnarray}
The full rapidity interval $\D Y$ is divided into $M$ equal bins: 
$\D Y=M\delta y$; each $y_i$ is within the $\delta y$-range and 
$\aver{n_m}\equiv\overline{\rho}_m\delta y \equiv \int_{\d y} \r_1(y)dy$.
An example for $q=2$ is given in Fig.~1a.
                 
\begin{figure}
\begin{center}
\epsfig{file=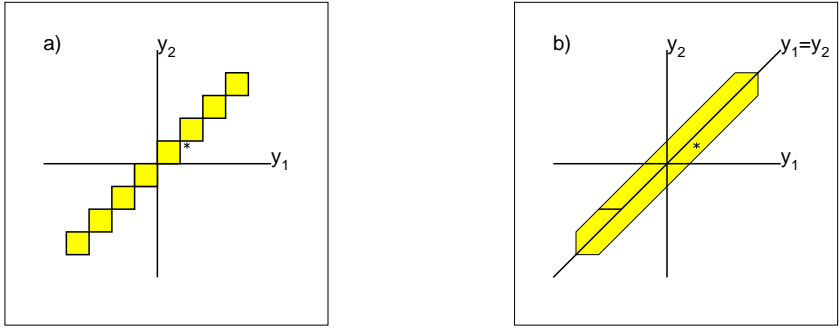,width=12cm}
\end{center}
\vs-4mm
\caption{Integration domains for a) the second-order factorial
moment and b) the second-order correlation integral. The asterisk in a) 
indicates the position of a particle pair with $|y_1-y_2|<\d y$ that is 
excluded from the $F_2$ calculation due to the binning. In b) the asterisk is 
included.}
\end{figure}

Likewise, cell-averaged normalized factorial cumulant moments may be defined as
\begin{equation}
K_q(\delta y) = \frac{1}{M(\delta y)^q}
 \sum^M_{m=1} \int\limits_{\d y} \frac{C_q(y_1,\ldots,y_q)\prod^q_{i=1} dy_i}
{(\bar\r_{m})^q} \ \ . \label{dr:47}
\end{equation}
They are related to the factorial moments by~\footnote{\ The higher-order 
relations can be found in~\cite{CaES91}.}
\begin{eqnarray}
F_2&=&1+K_2\ \ \ ,\nonumber\\
F_3&=&1+3K_2+K_3\ \ \ ,\nonumber\\
F_4&=&1+6K_2+3\overline{K^2_2}+4K_3+K_4 \,.\label{dr:48}
\end{eqnarray}
In $F_4$ and higher-order moments, ``bar averages" appear
defined as $\ol{AB}\equiv\sum\limits_m A_mB_m/M$.

To detect dynamical fluctuations in the density of particles produced in a
high-energy collision, a way must be devised to eliminate, or to reduce as
much as possible, the statistical fluctuations (noise) due to the finiteness
of the number of particles in the counting cell(s). This requirement can to a
large extent be satisfied by studying factorial moments and forms the basis 
of the factorial moment technique, known in optics, but rediscovered 
for multi-hadron physics in~\cite{bialas}. This crucial property does not
apply to, e.g., ordinary moments $\lan n^q\ran/\lan n\ran^q$.

The property of Poisson-noise suppression has made measurement of factorial
moments a standard technique, e.g. in quantum optics, to study the statistical
properties of arbitrary electromagnetic fields from photon-counting
distributions. Their utility was first explicitly recognized, for the single
time-interval case, in~\cite{Bedard:67:1} and later generalized to the 
multivariate case in~\cite{Cantrell:70}. 

\subsection{Density and Correlation Integrals}

A fruitful  recent development in the study of density fluctuations is 
the correlation strip-integral method.\cite{Hen83} By means of integrals 
of the inclusive density over a strip domain of Fig.~1b, rather than a 
sum of the box domains of Fig.~1a, one not only avoids unwanted side-effects 
such as splitting of density spikes, but also drastically increases the
integration volume (and therefore the statistical significance) at a given 
resolution.

Let us consider first the factorial moments $F_q$ defined according to 
Eq.~14. As shown in Fig.~1a for $q=2$,
the integration domain $\W_B=\sum^M_{m=1}\W_m$ consists of $M$
$q$-dimensional boxes $\W_m$ of edge length $\d y$.
A point in the $m$-th box corresponds to a pair $(y_1,y_2)$ of
distance $|y_1-y_2|<\d y$ and both particles in the same bin $m$.
Points with $|y_1-y_2|<\d y$ which happen {\it not} to lie in the same,
but in adjacent, bins (e.g., the asterisk in Fig.~1a) are left out.
The statistics can be approximately doubled by a change
of the integration volume $\W_B$ to  the strip-domain of Fig.~1b. 
For $q>2$, the increase of
integration volume (and reduction of squared statistical error)
is in fact roughly proportional to the order of the correlation.
The gain is even larger when working in two- or three-dimensional 
phase-space variables.

In terms of the strips (or hyper-tubes for $q>2$), the density 
integrals become
\beq
F^\rS_q(\d y) \equiv \frac{\int_{\W_s}\r_q(y_1\dots y_q){\prod}_i \rd y_i }
{\int_{\W_s}\r_1(y_1)\dots \r_1(y_q){\prod}_i \rd y_i }\ \ .
\label{3.45}
\eeq
These integrals can be evaluated directly from the data after selection
of a proper distance measure $(|y_i-y_j|,[(y_i-y_j)^2+(\f_i-\f_j)^2]^{1/2}$,
or better yet, the four-momentum difference $Q^2_{ij} = -(p_i-p_j)^2$)
and after definition of a proper multiparticle topology (GHP 
integral,\cite{Hen83} snake integral,\cite{CaSa89} star integral \cite{Star}).
Similarly, {\it correlation} integrals can be defined by replacing the
density $\r_q$ in Eq.~18 by the correlation function $C_q$.

The numerator 
of the factorial moments $F_q$ can be determined by counting, for 
each event, the number of $q$-tuples that have a pairwise $Q^2_{ij}$ 
smaller than a given value $Q^2$ and then averaging over all events. 
Using the Heaviside unit-step function ${\O}$, this can be mathematically 
expressed as
\beq
F^\rS_q(Q^2)=\frac{1}{\mbox{norm}} \left\lan q! \sum_{i_1<\dots<i_q} 
\prod_{all\ pairs\atop
k_1,k_2} {\O}(Q^2-Q^2_{i_{k_1},i_{k_2}})\right\ran\ \ ,
\eeq
where the factor $q$! takes into account the number of permutations
within a $q$-tuple. 
 
The normalization is  obtained from "mixed" events constructed by
random selection of tracks from different events in a
track pool. The multiplicity of a mixed event is taken to be a Poissonian
random variable, thereby ensuring that no extra correlations are introduced.
A correction factor is applied for the difference in average multiplicity of 
the Poissonian and the experimental distribution. The mixed events are 
treated in the same way as real events. 

\subsection{Power-Law Scaling}

The technique proposed in~\cite{bialas} consists of measuring the 
dependence of the normalized factorial moments (or correlation integrals) 
$F_q(\d y)$ as a function of the resolution $\d y$. For definiteness, 
$\delta y$ is supposed to be an interval in rapidity, but the method 
generalizes to arbitrary phase-space dimension, as occurs with the use 
of $Q^2$.

As pointed out above, the scaled factorial moments enjoy the property 
of ``noise-suppression". High-order moments further act as a filter 
and resolve the large $n_m$ tail of the multiplicity distribution.
They are thus  particularly sensitive to large density fluctuations 
at the various scales $\d y$ used in the analysis.

As proven in~\cite{bialas}, a ``smooth" (rapidity) distribution, which does 
not show any fluctuations except for the statistical ones, has the property 
of $F_q(\d y)$ being independent of the resolution $\d y$ in the limit 
$\d y\to  0$. On the other hand, if dynamical fluctuations exist and $P_\r$ 
is ``intermittent'', the $F_q$ obey the power law
\beq
F_q(\d y) \propto (\d y)^{-\f_q}\ , \ \ (\d y\to 0).
\eeq

The powers $\f_q$ (slopes in a double-log plot) are related~\cite{LiBu89} 
to the anomalous dimensions 
\beq
d_q=\f_q/(q-1)\ \ ,
\eeq
a measure for the 
deviation from an integer dimension. Equation 20 is a scaling law
since the ratio of the factorial moments at resolutions $L$ and $\ell$
\beq
R = F_q(\ell)/F_q(L) = (L/\ell)^{\f_q}
\eeq
only depends on  the ratio $L/\ell$, but not on $L$ and $\ell$,
themselves.

\begin{figure}[t]
\begin{minipage}[t]{5.9cm}
\hskip-0.3cm
\epsfig{file=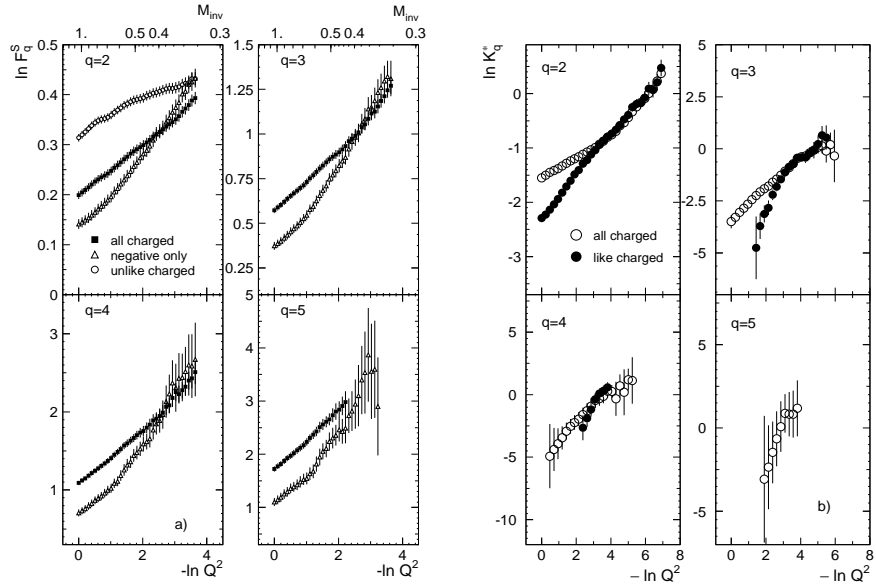,height=8.2cm,width=6cm} 
\end{minipage}
\begin{minipage}[t]{5.5cm}
\epsfig{file=fig2bsanta.ps,height=8.2cm,width=6cm} 
\end{minipage}\vskip-5mm
\caption{Correlation-integral method applied to NA22 hadron-hadron data 
\protect\cite{Agab93,NA22} in terms of a) ln $F^\rS_q$ and 
b) $\ln K^*_q$ as a function of $-\ln Q^2$.} 
\end{figure}

As pointed out above, the experimental  study of  correlations is already
difficult for three particles. The close connection between correlations and 
factorial moments offers the possibility of measuring higher-order correlations
with the factorial moment technique at smaller distances than was previously 
feasible. Via Eq.~21, the method further relates possible scaling behavior 
of such correlations to the physics of fractal objects.

One further has to  stress the advantages of factorial cumulants 
compared to factorial moments, since the former measure {\it genuine}
correlation patterns, whereas the latter contain additional large 
combinatorial terms which mask the underlying dynamical correlations.

The definition of ``intermittency" given in (20), has its origin in other 
disciplines.\footnote{\ For a masterly expos{\'e} of this subject 
see~\cite{Zeld90}.} It rests on a loose parallel between the high 
non-uniformity of the distribution of energy dissipation, for example, 
in turbulent intermittency and the occurrence of large ``spikes" in 
hadronic multiparticle final states. In the following we use the term 
``intermittency'' in a weaker sense, meaning the rise of factorial moments 
with increasing resolution, not necessarily according to a strict power law.

\section{The State of the Art}

The suggestion that normalized factorial moments of particle distributions 
might show power-law behavior has spurred a vigorous experimental search 
for (more or less) linear dependence of $\ln F_q$ on $-\ln\d y$. A full
review of the present situation is given in \cite{Wolf93}.

\begin{figure}[tH]
\begin{center}
%{\sf DIFFERENTIAL CORRELATION INTEGRALS (Q$^2$)}\vs3mm
\hs-8cm \begin{sideways} $\Delta F_2$\end{sideways}
\vs -0.8cm
%\vskip-0.2cm
\epsfig{file=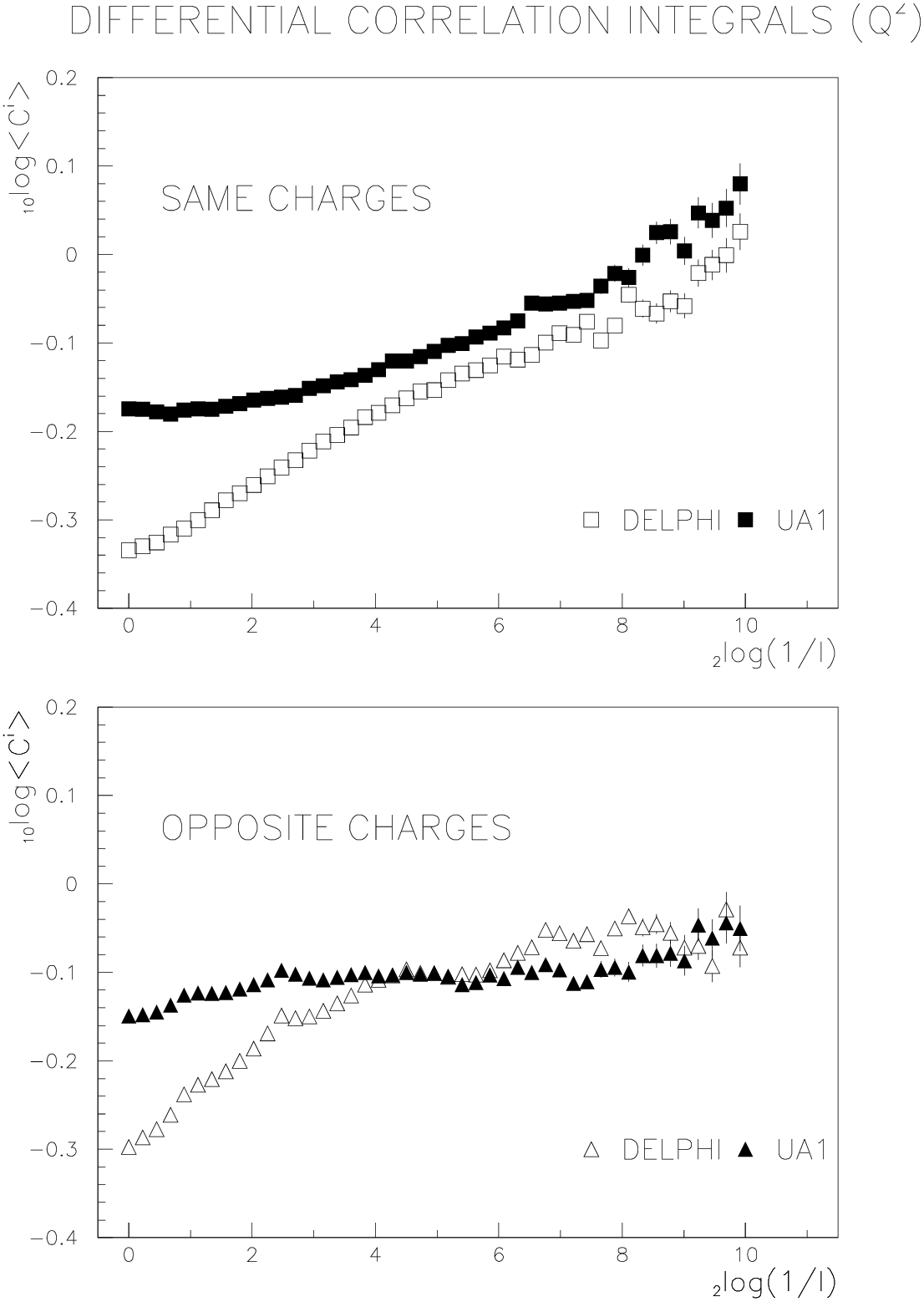,bbllx=50pt,bblly=40pt,bburx=580pt,bbury=772pt,clip=,
width=6.7cm}
\end{center}
\vs-1mm
\caption{Comparison of density integrals for $q=2$ in
their differential form $\Delta F_2$ (in intervals $Q^2, Q^2+\rd Q^2$) 
as a function of $_2\log(1/Q^2)$ for e$^+$e$^-$ (DELPHI, preliminary) 
and hadron-hadron collisions (UA1).\protect\cite{Abreu94}}
%\end{minipage}
\end{figure}

As an example, we give Fig.~2a where hadron-hadron data~\cite{Agab93} 
on $F^\rS_q$ are 
plotted as a function of $-\ln Q^2$, with all two-particle combinations in an 
$n$-tuple having $Q^2_{ij}<Q^2$. The
following observations can be made:\\
i)\ \ \  the moments show a  steep rise with decreasing $Q^2$;\\
ii)\ \  negatives are much steeper than all-charged, \\
iii)\  $F^\rS_2$ is flatter for $(+-)$ than for all-charged or like-charged
combinations.
             
The last two observations directly demonstrate the large influence 
of {\it identical} particle correlations on the factorial moments and
their scaling behavior. These results agree very well with 
results from the UA1 collaboration.\cite{Neum93} In Fig.~2b it is,
furthermore, shown~\cite{NA22} in terms of the $K^*_q(Q^2)$ (using
the star topology) that genuine correlations exist and increase 
with improving resolution at least up to order $q=5$ in hadron-hadron  
collisions. Again, like-charged particle combinations increase faster than
all-charged ones.

Of particular interest is a comparison of hadron-hadron to e$^+$e$^-$ results 
in terms of same and opposite charges of the particles involved. This 
comparison has been done for UA1 and DELPHI data in~\cite{Abreu94} and is 
shown in Fig.~3 for $q=2$ (in fact, in this figure a differential form 
$\Delta F_2$ of Eq.~19 is presented). An important difference between UA1 
and DELPHI can be observed in a comparison of the two sub-figures: 

For relatively large $Q^2(>0.03$ GeV$^2$), where Bose-Einstein effects do 
not play a major role, the e$^+$e$^-$ data increase much faster with 
increasing $-_2\log Q^2$ than the hadron-hadron results. For e$^+$e$^-$, 
the increase in this $Q^2$ region is very similar for same and for 
opposite-sign charges. 

At small $Q^2$, however, the e$^+$e$^-$ results approach the hadron-hadron 
results. 

The authors conclude that for e$^+$e$^-$ annihilation at LEP at least
two processes are responsible for the power-law behavior: Bose-Einstein
correlation following the evolution of jets. In  hadron-hadron collisions 
at present collider energies, the Bose-Einstein effects seem 
mostly relevant. 
                  
The exact functional form of $\Delta F_2$ or $\Delta K_2$ is derived from 
the data of UA1 \cite{Neum93,Eggers} and NA22,\cite{Agab93} again in its 
differential form, in Fig.~4. Clearly, the low-$Q$ data favor a power law 
in $Q$ over a (non-scaling) exponential, double-exponential or Gaussian law. 

In Fig.~5, the NA22 results for two-, three- and four-particle Bose-Einstein
correlations (equivalent to $\Delta F_2(Q^2)$ are compared to a 
multi-Gaussian para\-metrization.\cite{Bial92} In the conventional linear 
presentation of Fig.~5a the fits look reasonable. If the same data and same 
fits are presented in a log-log scale, however, the power-law deviation 
from the multi-Gaussians starts to become visible (Fig.~5b).

\begin{figure}
\vskip-0.8cm
\begin{tabular}{ll}
\epsfig{file=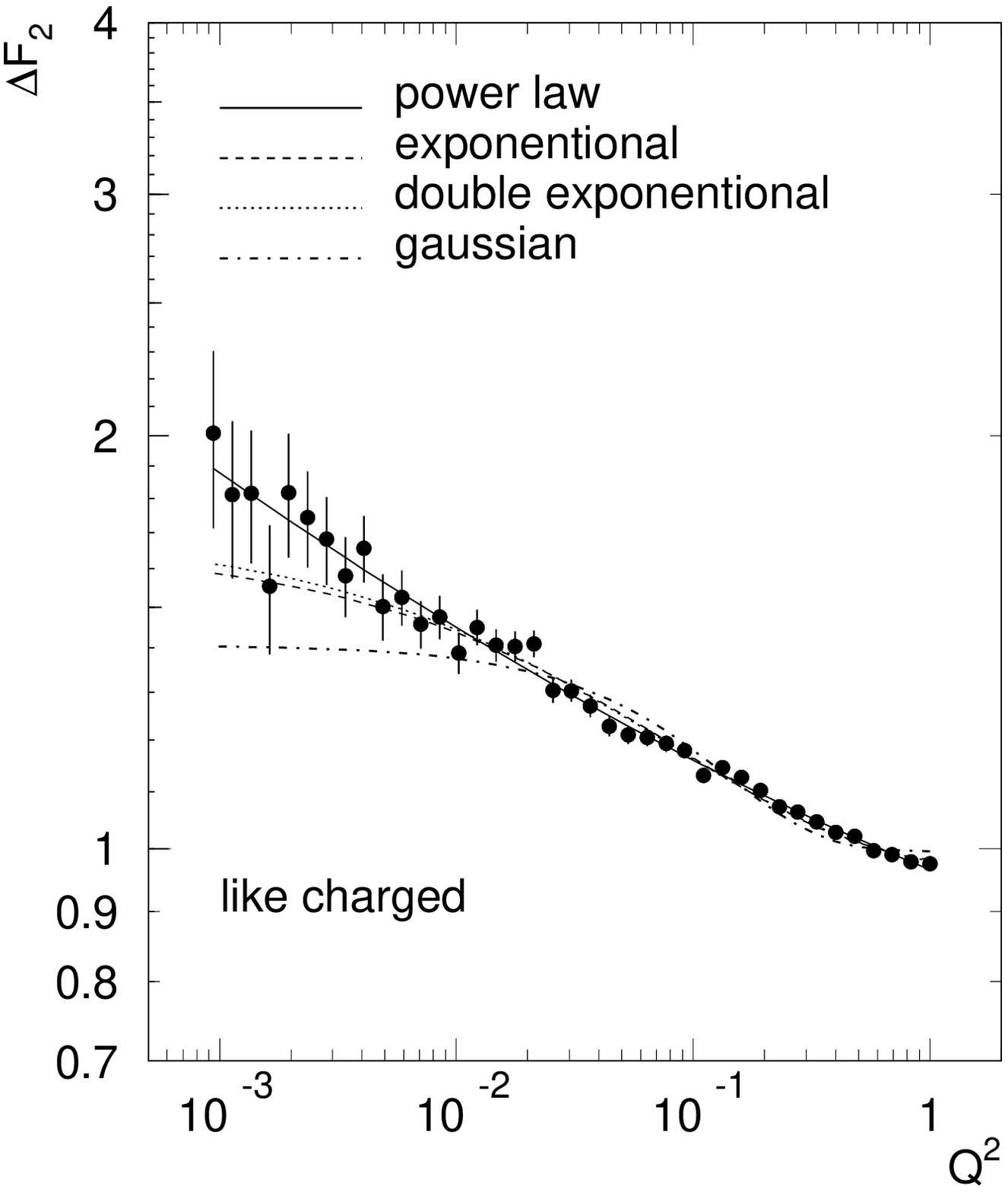,width=5.4cm} &
\epsfig{file=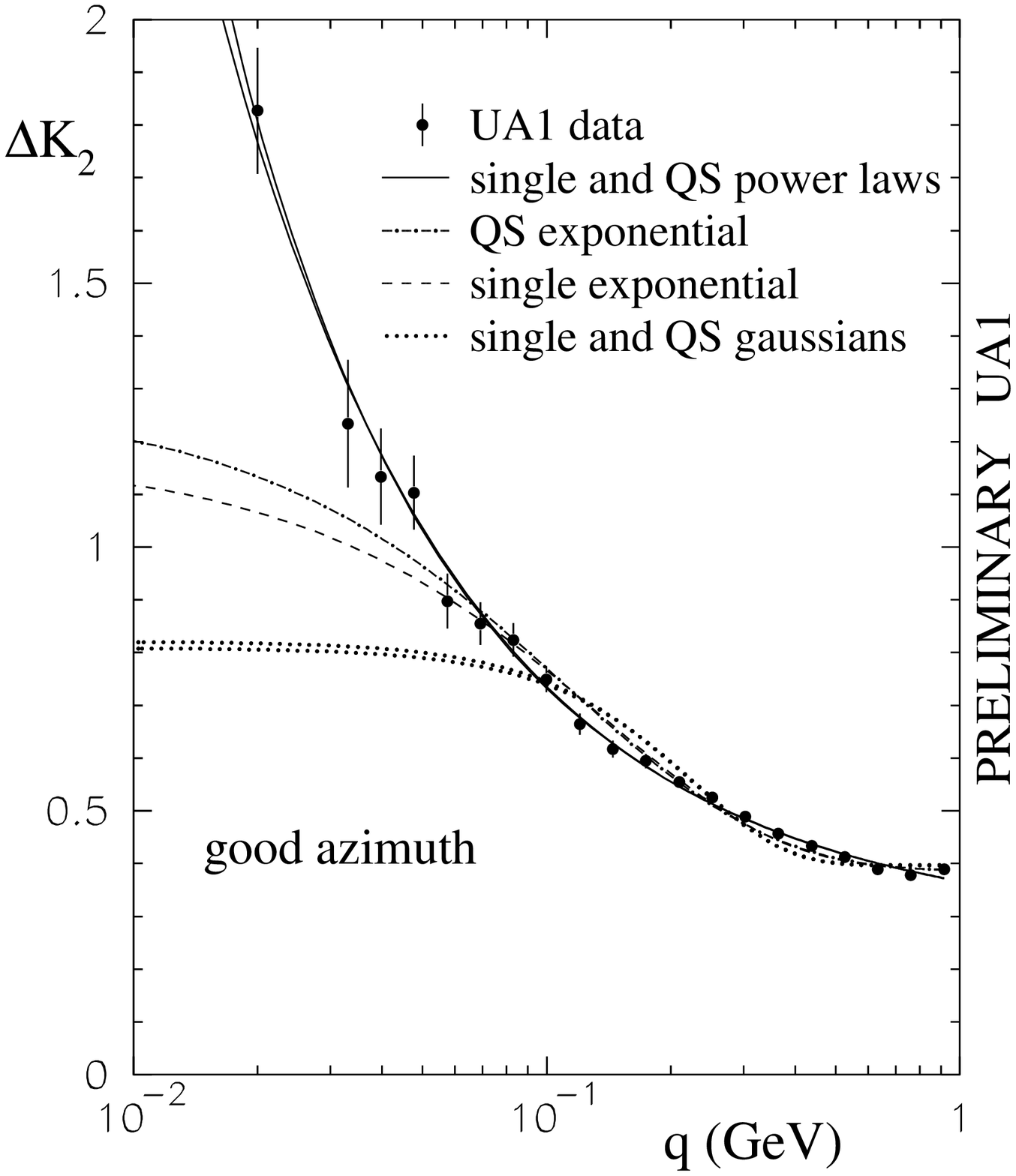,width=5.5cm} 
\end{tabular}

\vs-3mm
\caption{Density integral $\Delta F_2$ as a function of log 
$Q^2$ for NA22~\protect\cite{Agab93}, correlation integral $\Delta K_2$ 
as a function of log $Q$ for like-charged pairs in UA1 (preliminary)
\protect\cite{Eggers} compared to power-law, exponential, 
double-exponential and Gaussian fits, as indicated.}
\vs-3mm
\begin{center}
\epsfig{file=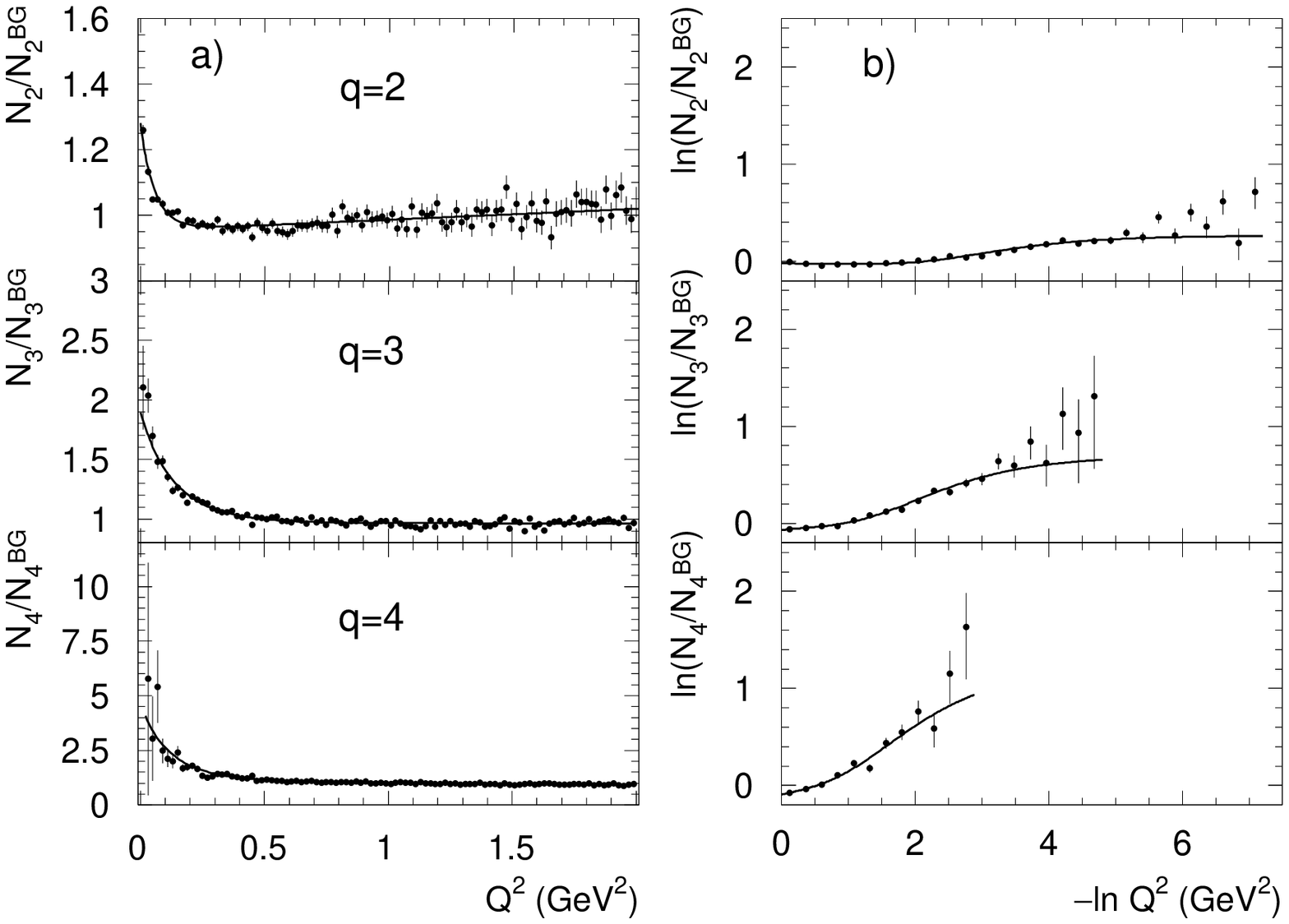,width=11cm} 
\end{center}
\vs-3mm
\caption{The normalized two-, three- and four-particle inclusive 
densities (equivalent to $\Delta F_q$) as a function of $Q^2$ (left) 
and $-\ln Q^2$ (right).\protect\cite{Agab95} Curves show the 
multi-Gaussian fits according to \protect\cite{biya90}. }
\end{figure}

If the observed effect is real, it supports a view recently developed
in~\cite{Bial92}. In that paper, intermittency is explained from Bose-Einstein
correlations between (like-sign) pions. As such, Bose-Einstein 
correlations from a static source are not power-behaved. A power law is 
obtained if i) the size of the interaction region is allowed to fluctuate, 
and/or ii) the interaction region itself is assumed to be a self-similar 
object extending over a large volume. Condition ii) would be realized if 
parton avalanches were to arrange themselves into  self-organized critical 
states.\cite{Bak87} Though quite speculative at this moment, it is an 
interesting new idea with possibly far-reaching implications. 
We should mention also that in such a  scheme 
intermittency is viewed as a final-state interaction effect and
is therefore not troubled by hadronization effects.  
 
In perturbative QCD, on the other hand, the fractal structure of jets
follows, in principle, from parton branching and the intermittency indices 
$\f_q$ are directly
related to the anomalous multiplicity dimension $\g_0 = (6\a_s/\p)^{1/2}$
\cite{Gust91,Ochs92,Doks93,Brax94} and, therefore, to the running 
coupling constant $\a_s$ via the simple relation,
$$ \f_q=(q-1)D-\frac{q^2-1}{q} \gamma_0\ ,$$
where $D$ is the usual topological dimension of the analysis. 
In the same theoretical context, it has been 
argued \cite{Ochs92,Doks93,Brax94} that the 
opening angle between particles is a suitable and sensitive variable
to analyze and well suited for these first ana\-ly\-ti\-cal QCD 
cal\-cu\-la\-tions
of higher-order correlations. It is, of course, closely related to $Q^2$.

A first analytical QCD calculation~\cite{Ochs92} is based on the 
double-log approximation (DLA) with angular ordering~\cite{ao} (for a
recent experimental study of angular ordering see~\cite{Acci95})
and on local parton-hadron duality.\cite{lphd} A preliminary 
comparison with DELPHI data~\cite{Mandl94} gives encouraging results, 
but shows that the situation is far from trivial, since finite-energy
effects, four-momentum cut-offs, resonance decays etc. still dominate at
LEP energies.

\section{Summary}

Multiparticle production in high-energy collisions is an ideal field
to study genuine higher-order correlations. Methods also used in other
fields are being tested and extended here for general application.
Indications for genuine, approximately self-similar higher-order
correlations are indeed found in hadron-hadron collisions, but still 
need to be establised in their genuine and self-similar character in
e$^+$e$^-$ collisions at high energies.  At large four-momentum
distance $Q^2$, they are not only expected to be an inherent property
of perturbative QCD, but are directly related to the anomalous
multiplicity dimension and, therefore, to the running coupling
constant $\a_s$. At small $Q^2$, the QCD effects are complemented by
Bose-Einstein interference of identical mesons carrying information on
the unknown space-time development of particle production during the
collision. The interplay between these two mechanisms, particularly
important for an understanding of the process of hadronization, is
completely unknown at the moment.

\end{document}